# Imaging Quantum Spin Hall Edges in Monolayer WTe$_2$

Yanmeng Shi[1]*, Joshua Kahn[2]*, Ben Niu[1,3]*, Zaiyao Fei[2], Bosong Sun[2], Xinghan Cai[2], Brian A. Francisco[1], Di Wu[3], Zhi-Xun Shen[4], Xiaodong Xu[2†], David H. Cobden[2†], Yong-Tao Cui[1†]

1. Department of Physics and Astronomy, University of California, Riverside, CA 92521, USA.
2. Department of Physics, University of Washington, Seattle, Washington 98195, USA.
3. National Laboratory of Solid State Microstructures, Department of Materials Science and Engineering, College of Engineering and Applied Sciences, and Collaborative Innovation Center of Advanced Microstructures, Nanjing University, Nanjing 210093, China.
4. Geballe Laboratory for Advanced Materials, Stanford University, Stanford, CA 94305, USA.

* These authors contribute equally to this work.
† Email: xuxd@uw.edu; cobden@uw.edu; yongtao.cui@ucr.edu

A two-dimensional (2D) topological insulator (TI) exhibits the quantum spin Hall (QSH) effect, in which topologically protected spin-polarized conducting channels exist at the sample edges[1–6]. Experimental signatures of the QSH effect have recently been reported for the first time in an atomically thin material, monolayer WTe$_2$. Electrical transport measurements on exfoliated samples[7,8] and scanning tunneling spectroscopy on epitaxially grown monolayer islands[9,10] signal the existence of edge modes with conductance approaching the quantized value. Here, we directly image the local conductivity of monolayer WTe$_2$ devices using microwave impedance microscopy, establishing beyond doubt that conduction is indeed strongly localized to the physical edges at temperatures up to 77 K and above. The edge conductivity shows no gap as a function of gate voltage, ruling out trivial conduction due to band bending or in-gap states, and is suppressed by magnetic field as expected. Interestingly, we observe additional conducting lines and rings within most samples which can be explained by edge states following boundaries between topologically trivial and non-trivial regions. These observations will be critical for interpreting and improving the properties of devices incorporating WTe$_2$ or other air-sensitive 2D materials. At the same time, they reveal the robustness of the QSH channels and the potential to engineer and pattern them by chemical or mechanical means in the monolayer material platform.

In a 2D TI, gapless states are guaranteed to exist at the edges separating the topologically non-trivial bulk from the topologically trivial surroundings[1–5]. When the chemical potential lies in the bulk energy gap, charge transport can occur only through these edge states. Because the edge states are helical (the electron spin is locked to its momentum), elastic backscattering is suppressed due to time reversal symmetry and under ideal circumstances this gives ballistic electron transport and consequently a quantized conductance, suppressed by magnetic field, which is considered the hallmark of the QSH effect[11].

Signatures of the QSH effect have been reported in HgTe/CdHgTe quantum wells[12–15] and InAs/GaSb double quantum wells[16–19], and more recently in monolayers of the layered semimetal WTe$_2$[7–10]. These experiments, however, presented some puzzles. Although edge conduction was seen in all these systems, the measured conductance often showed large deviations from the expected quantized value[12–14,16,17,19]. Also, a combination of transport and scanning probe microscopy studies on the quantum well systems revealed that topologically trivial conduction could also occur at the edges, and that the magnetic field dependence was surprisingly small[20–22]. It should thus be revealing to examine this new monolayer system in the same way. Here, we apply microwave impedance microscopy (MIM)[22–25] to monolayer WTe$_2$ devices, directly mapping the local conductivity. We observe gapless conduction localized to the sample edges, whose electrical and geometrical behavior is consistent with the QSH effect. In addition, MIM reveals other important aspects of the conduction properties of this system that could not be deduced from transport, providing crucial insights for optimizing devices and opportunities to access and manipulate the helical edge channels.

The MIM technique probes the local conductivity by analyzing the imaginary and real parts of the complex admittance, which we call MIM-Im and MIM-Re respectively, between a sharp conducting tip and the sample (Fig. 1a). MIM-Im characterizes the amount of screening of the microwave electric field at the tip by the sample, while MIM-Re characterizes the dissipation generated by the induced oscillating currents in the sample. The responses of MIM-Im and MIM-Re to changing local resistivity can be obtained via finite element analysis (an example is plotted in Fig. 2e): MIM-Im increases monotonically as the resistivity decreases, while MIM-Re is strongly peaked at an intermediate resistivity value. These response curves serve as guides for interpreting the MIM measurements.

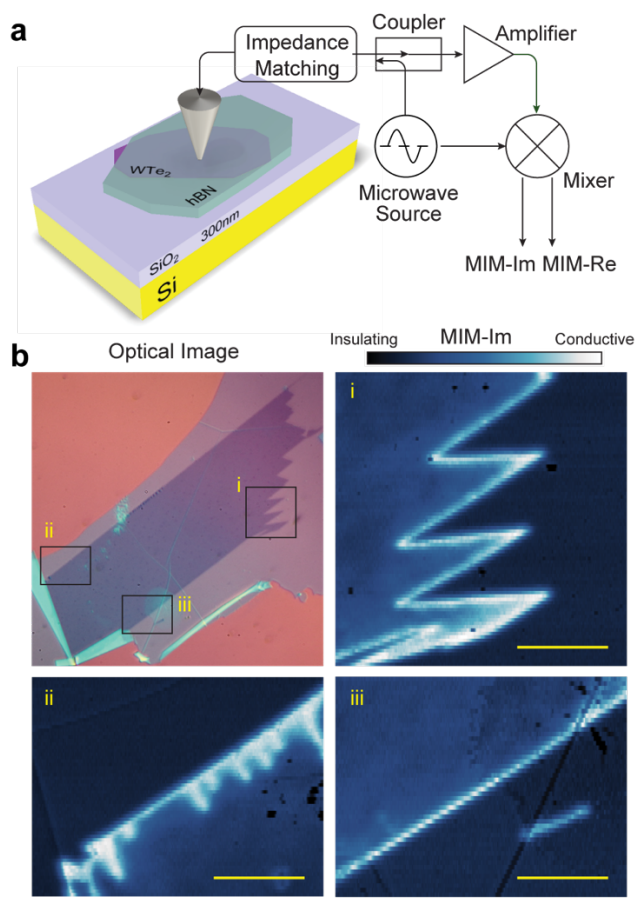

**Figure 1 | Imaging edge conductivity in monolayer WTe$_2$. a**, Schematics of the technique and device structure. **b**, Optical image (top left) of an WTe$_2$ monolayer exfoliated onto SiO$_2$ and covered by a 10-nm thick hBN, together with MIM-Im images of the regions marked i, ii, and iii, measured at $T$=8 K and $B$=0. All scale bars are 5 μm.

We first demonstrate the edge conduction in a simple sample structure: a monolayer WTe$_2$ flake exfoliated onto SiO$_2$ and directly covered with a larger flake of hexagonal boron nitride (hBN) (See Methods for fabrication details and Supplementary Table S1 for information of all samples).

In Fig. 1b an optical image is shown at the top left, and the adjacent images are MIM-Im measurements of selected regions, labelled i-iii, taken at temperature $T$=8 K and magnetic field $B$=0. The MIM-Im signal in the interior of the flake is comparable to that over the substrate, indicating a highly insulating state. Meanwhile, a bright narrow line indicating much higher conductivity follows the entire outline of the flake, whether it is straight (region iii) or sharply zigzagged (region i). Moreover, in region ii we see a series of short bright lines intruding into the interior that do not correspond to features visible optically. These are almost certainly due to small tears which are narrow enough that the two opposing edges of each tear cannot be resolved; they thus appear as single lines, albeit thicker and brighter than the line following the outer edge. These observations unambiguously confirm the existence of edge conduction in monolayer $WTe_2$. Furthermore, the strict conformity with the microscopic edge geometry agrees with the topological nature of helical edge channels (see Supplementary SI-2).

In a 2D TI the helical edge state dispersion spans the bulk gap, connecting the conduction and valence bands, so the edge conduction should persist while the bulk chemical potential is tuned across the gap[11,26]. This property is in contrast with edge conduction due to in-gap states or to band bending in a trivial insulator, which exist only over a limited energy range[27–29]. To study this behavior, we turn to samples with electrical contacts which allow simultaneous MIM and transport measurements, while also applying a gate voltage $V_g$ to the silicon substrate (see Supplementary SI-3). Figure 2 presents results from a monolayer $WTe_2$ device incorporating two thin graphite contacts separated by ~1.2 μm. Fig. 2a shows MIM-Im and MIM-Re images at $B$=0 and $V_g$=-15 V, which we identify as the charge neutrality point (CNP) in the bulk through gate voltage dependence measurement presented below. Again, the small MIM signals in the interior of the flake signify a highly insulating state, while the large MIM-Im signal near the $WTe_2$ edge indicates a highly conductive state.

To probe the gate voltage dependence of both the edge and bulk conductivity, the tip is scanned repeatedly along a single line crossing the $WTe_2$ edge halfway between the contacts as $V_g$ is varied from -60 V to +40 V (Fig. 2b). Over this range the bulk goes from mildly conducting p-doped through insulating to highly conducting n-doped, implying that the chemical potential is tuned across the bulk gap. All the while, the edge remains highly conductive, just as expected for a topological edge mode. Near the CNP, since both MIM-Im and MIM-Re in the bulk are small, the conductivity value falls on the highly insulating side of the response curve (Fig. 2e). Upon gating to either n- or p-doped side the bulk conductivity increases, so MIM-Im should increase monotonically while MIM-Re should exhibit a peak. The measured bulk signals show exactly this behavior (see the annotations in Figs. 2b and 2e), albeit with an asymmetry suggesting different carrier mobilities on the p-doped and n-doped sides.

In a magnetic field the helical edge states are expected to mix, opening a Zeeman gap and suppressing conduction when the chemical potential at the edge is near this gap[26]. In previous dc transport measurements, the edge conduction was indeed suppressed, although with a complicated dependence on gate voltage probably due to disorder effects[7,8]. Figure 2c shows MIM images like those in Fig. 2a but taken in a perpendicular field $B$=9 T. Edge conduction can still be clearly resolved. There is a visible increase in MIM-Re when the field is applied, which occurs consistently over a wide range of doping levels as seen in the gate dependence in Fig. 2d. On the other hand, the MIM signals in the bulk are little changed, indicating that the effect of magnetic field is primarily on the edges, consistent with transport measurements[7,8]. To study the effect on the edge signals more quantitatively, we plot in Fig. 2f MIM traces taken across the edge at 0 T and 9 T on the same axes. To improve signal-to-noise ratios these traces are obtained by averaging horizontal linecuts in the MIM images taken near the CNP with the edge position aligned. We see that when 9 T is applied MIM-Im decreases while MIM-Re increases. Referring to the MIM response curves, these changes indicate an increase in the edge resistivity (see the dotted lines in Fig. 2e) that is consistent with the dc transport results[7,8]. (See Supplementary SI-4 for the simulation.)

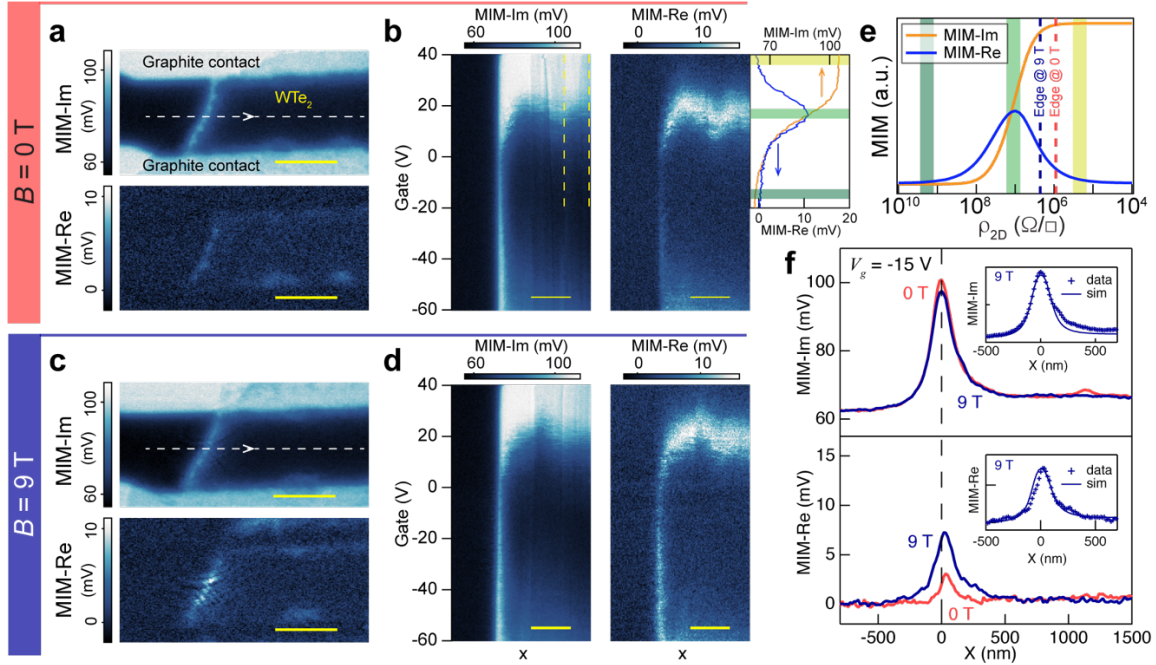

**Figure 2 | Gate and magnetic field dependence of the edge conduction. a**, MIM-Im and MIM-Re images of part of a monolayer WTe$_2$ flake between two thin graphite contacts, measured at $B$=0. **b**, MIM-Im and MIM-Re obtained along the dashed line in (a) and stacked as a function of gate voltage. The upper right panel plots the averaged line cuts over the bulk region indicated by the two dashed lines in the MIM-Im channel, from $V_g$=-20 V to 40 V. **c** and **d**, Real-space images and gate voltage dependence at $B$=9 T. **e**, MIM-Im and MIM-Re signals as a function of 2D resistivity. The colored bands match those in the line cuts in (b), and the dotted lines indicate the 2D resistivity at the edge for B=0 and 9 T, estimated from the line traces in (f). **f**, Averaged MIM-Im (top) and MIM-Re (bottom) traces of linecuts in between the two contacts with the edge position aligned, at both $B$=0 and 9 T. Insets are the comparison between the 9 T line traces and the simulated MIM responses. The measurement temperature is 5 K. All scale bars are 1 μm.

As well as at the physical edges, in the interior of samples conducting features are frequently observed that can be interpreted as boundary modes separating regions of different topological characters, as illustrated in Figs. 3 and 4. Figure 3a is an optical image of a monolayer WTe$_2$ flake on SiO$_2$ that is mostly covered by an hBN flake, though a corner is exposed and there is also a folded bilayer region. In the MIM-Im image (Fig. 3b), a conducting line is seen all along the exterior monolayer edge where it is under the hBN, but not on the edge of the bilayer. This observation is consistent with the bilayer not supporting edge states, as reported before[7]. Also, at the exposed part the conducting line runs along the edge of the hBN rather than around the exterior edge. Since the exposed part will be oxidized and thus trivially insulating, this behavior is just as expected given the topological nature of the QSH edge states. In addition, we see conducting lines cutting across the interior of the monolayer under the hBN. These lines are not likely to be folds because they are not visible in topography, as measured by atomic force microscopy (AFM) (see Supplementary SI-5). We can also rule out grain boundaries using polarized Raman spectroscopy: the 1T' structure is orthorhombic, so the intensities of all Raman peaks show a two-fold rotationally symmetric dependence on the polarization angle relative to the crystal axes (Fig. 3c). Figure 3d is a polar plot of the Raman intensity at each of the five points labeled in Fig. 3b, showing that the crystal axes are everywhere the same. The only explanation remaining is narrow cracks, generated during either exfoliation of the WTe$_2$ or transfer of the hBN on top. QSH edge states should follow both edges of a crack, and indeed the gate dependence of the internal lines is very similar to that of the exterior edges (see Supplementary SI-6). Figures 3e and 3f are MIM-Im images of another sample showing that the edge conduction, including that at internal cracks, dominates the conductivity even at 77 K and is still visible up to 100 K, again consistent with transport measurements[7,8].

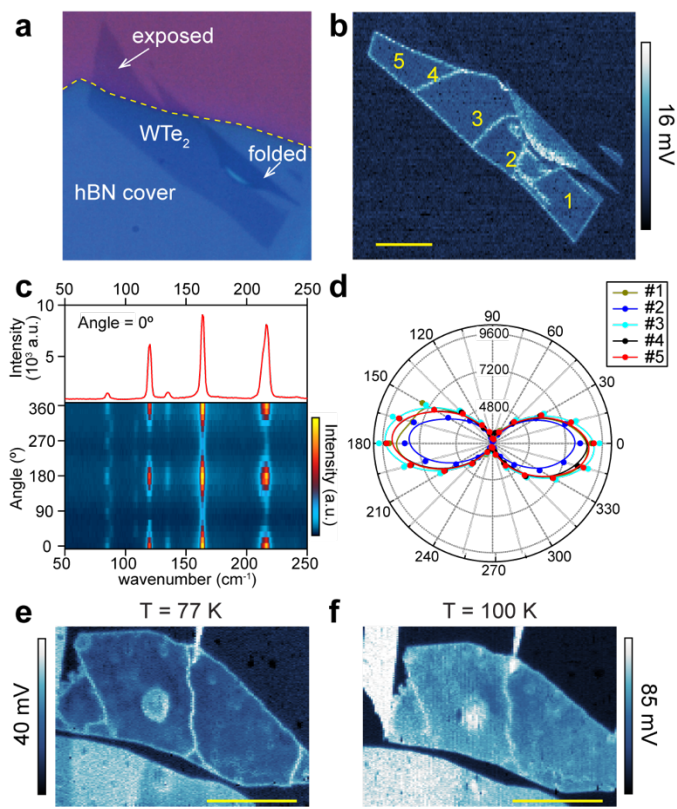

**Figure 3 | Conduction at oxidized edges and internal cracks in monolayer WTe$_2$. a**, Optical image of a monolayer WTe$_2$ flake partially covered by hBN. **b**, MIM image of the same flake measured at $T$=480 mK and $B$=0. High conductivity is observed both at the physical edges and along lines in the interior. **c**, Polarized Raman spectroscopy and angular dependence measured at 5 K, for spot #5 marked in (b). **d**, Polar plot of the 163 cm$^{-1}$ Raman peak intensity for the five spots marked in (b). All have the same angular dependence, showing that the crystal axes are the same and implying that the lines are cracks. **e** and **f**, MIM-Im images of another monolayer WTe$_2$ sample at 77 K (e) and 100 K (f). All scale bars are 5 µm.

In Fig. 4 we illustrate some other phenomena revealed by our technique which have important consequences for device fabrication and performance. The top panels in Figs. 4a-c show MIM-Im images of regions of three different monolayer WTe$_2$ devices, each with a pair of encapsulated thin Pt contacts and each showing poor dc electrical characteristics. All exhibit not only conducting lines at the edges but also conductivity elsewhere in various patterns. In Fig. 4a one can discern a strip of enhanced conductivity in the WTe$_2$ bulk adjacent to the contact edges, most apparent in the MIM-Re image shown in the lower panel. Similar features were seen in other devices, including the one with graphite contacts presented in Fig. 2. On the other hand, in Fig. 4b a narrow dark strip of very low conductivity surrounds each contact, separated by a bright conducting line from the monolayer bulk. In Figs. 4b and 4c we see a network of conducting internal lines similar to those in Fig. 3b. In addition, in Fig. 4c we also see many small conducting rings, which match well the outlines of small blisters of height 3 - 5 nm visible in the simultaneous topography scan (lower panel). Two similar but fainter rings are just discernable in Fig. 4a.

The cartoon in Fig. 4d illustrates our interpretation of these features. First, there are cracks and tears which the edge modes conform to, as described earlier. Next, the strips around the contacts can be explained by oxidation due to "tenting" of the hBN/$WTe_2$ over the edges of the contact metal, which permits access by air or liquids. Weak oxidation increases conductivity; stronger oxidation produces a trivial insulator, leading to an edge mode at its boundary with the unoxidized $WTe_2$ as seen in Fig. 3b. The blisters in Fig. 4c are probably also a result of chemical damage that turns spots of the $WTe_2$ into a different insulating material, leaving a ring-shaped edge mode surrounding each internal (topologically trivial) hole in the QSH material. Another possibility that we considered is that strain around the blisters modifies the electronic structure[6], but the blisters are much shallower than those typically produced by trapped contamination[30] and are unlikely to produce the required ring of high strain.

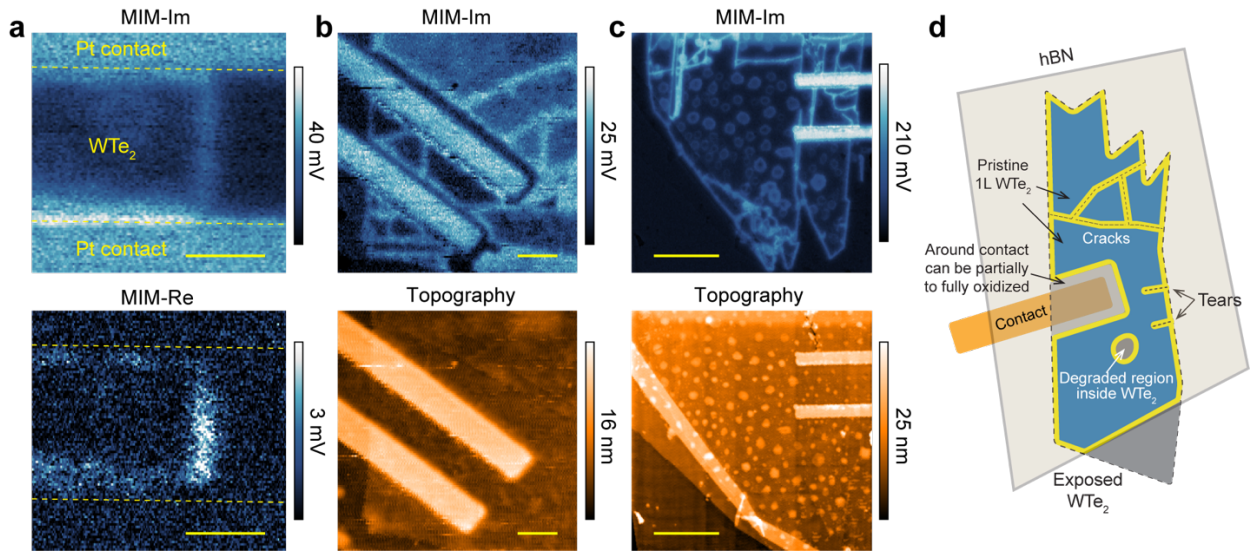

**Figure 4 | Conductivity features near contacts and around defects in monolayer $WTe_2$. a**, MIM-Im and MIM-Re images of part of a monolayer $WTe_2$ device between two Pt contacts, measured at $T$=480 mK, $V_g$=0 V, and $B$=12 T. Scale bars are 500 nm. **b**, MIM-Im and topography images for part of a second monolayer $WTe_2$ device, measured at $T$=10 K, $V_g$=3.3 V, and $B$=0 T. In the topography the flake appears continuous, but MIM reveals that the regions around the contacts are highly insulating (dark). Scale bars are 1 μm. **c**, MIM-Im and topography images for part of a third monolayer $WTe_2$ device, measured at $T$=10 K, $V_g$=0 V, and $B$=9 T. The small rings visible in the MIM-Im image correspond to the blisters in the topography image. Scale bars are 3 μm. **d**, Cartoon illustrating various conductivity features observed in our experiments.

The presence of such cracks, rings, contact-surrounding strips, and other internal features seen in Figs. 3 and 4 could produce misleading dc transport results. For instance, they could conduct in parallel with the exterior edges, introduce large contact resistances, or block current along edges, depending on details. Their presence is difficult to detect and control during device fabrication. Fortunately, MIM can be performed on transport device structures (prior to placing a top gate) to probe them. In addition, our results imply that in this monolayer QSH system edge channels could be created on demand, for various applications, by locally inducing either oxidation or fracture. That could, for example, allow the helical modes on two edges to be brought very close to each other or to other objects such as superconductors, or allow modes to be crossed and tunnel-coupled with each other, which may offer new opportunities for the study and manipulation[31] of QSH edge modes. Such configurations represent an advantage of atomically thin materials as they are difficult to achieve in semiconductor heterostructure QSH systems where the edge states are buried and are defined by depletion fields[12,16].

## Methods

**Device Fabrication.** Graphite, hBN, and $WTe_2$ flakes were prepared by mechanical exfoliation of crystals onto a 285 nm $SiO_2$ substrate. The graphite and hBN were exfoliated in ambient conditions. The $WTe_2$ was exfoliated in an inert atmosphere, provided by a nitrogen-filled glove box, to avoid degradation. Suitable flakes were optically identified and a polymer stamp transfer method[32] was used to construct the devices. The simplest devices were made by placing an hBN onto a monolayer of $WTe_2$. This capping hBN allowed the $WTe_2$ to be removed from the glove box without oxidation. Devices with contacts also required a bottom layer of hBN in order to properly encapsulate the $WTe_2$. The contacts were either exfoliated graphite flakes or evaporated V/Au or Pt metal patterned using standard e-beam lithography and lift-off processes.

**Microwave Impedance Microscopy Measurement.** MIM measurements were performed by delivering a small microwave excitation of ~0.1 μW at a fixed frequency in the range 1 - 10 GHz to a chemically etched tungsten tip[25]. The reflected signal was analyzed to extract the demodulated output channels, MIM-Im and MIM-Re, which are proportional to the imaginary and real parts of the admittance between the tip and the sample, respectively. To enhance the MIM signal quality, the tip was excited to oscillate at a frequency of ~32 kHz with an amplitude of ~8 nm. The resulting oscillation amplitudes of MIM-Im and MIM-Re were then extracted using a lock-in amplifier to yield d(MIM-Im)/d$z$ and d(MIM-Re)/d$z$, respectively. The d(MIM)/d$z$ signals are free of fluctuating backgrounds, thus enabling more quantitative analysis, while their behavior is very similar to that of the standard MIM signals[25]. In this paper we simply refer to d(MIM)/d$z$ as the MIM signal, and the simulations were done accordingly.

**Raman Characterization.** The collinear-polarization low-frequency Raman spectroscopy was performed under normal incidence using a diode-pumped solid-state laser with an excitation wavelength of 532 nm. A linearly polarized laser beam was focused on the sample by a 40x objective to a diameter of 1~2 μm. The reflected radiation with polarization parallel to the inherent polarization of the excitation beam was collected by a grating spectrometer equipped with a thermoelectrically cooled charge-coupled detector (CCD). The Rayleigh line was suppressed using three notch filters with an optical density of 3 and a spectral bandwidth of ~10 $cm^{-1}$. A typical laser power of 0.25 mW was used to avoid sample heating. The sample temperature was fixed at 5 K.

**Acknowledgements**
The work at UCR was supported by the startup funds from the Regents of the University of California. The work at UW was supported by the US Department of Energy, Office of Basic Energy Sciences, Awards DE-SC0002197 (D.H.C.) and DE-SC0018171 (X.X.). Z.F. and X.C. were partially supported by NSF MRSEC 1719797. B.N. and D.W. acknowledge financial support from State Key Program for Basic Research of China (Contract no. 2015CB921203). B.N.'s visit at UCR was supported by China Scholarship Council (CSC).

**Author Contributions**
Y.T.C., D.H.C. and X.X. conceived the experiment. Y.S., B.N., B.A.F. and Y.T.C. performed the MIM measurements; J.K., Z.F. and B.S. fabricated the devices; X.C. performed the Raman measurements; Y.T.C. and B.N. performed the finite element simulation; Z.X.S. provided access to MIM measurements at He-3 temperature; Y.T.C., D.H.C., and X.X. analyzed the data and wrote the paper with comments from all authors.

**Competing financial interests**
The authors declare no competing financial interests.